\documentclass[12pt, authoryear]{article}

\usepackage[latin1]{inputenc}
\usepackage{graphicx, geometry,  amsmath, amsthm, amssymb,   fleqn, amsfonts,  subfigure, amstext}
\usepackage{wrapfig, amsopn, latexsym,  epsfig, multirow, multicol, verbatim}
\usepackage{comment, fancyhdr,  bm}

\usepackage[affil-it]{authblk}
\usepackage[english]{babel}

\title{Inference for three-parameter $M$-Wright distributions with applications}
 
\date{}

\author{Dexter O. Cahoy, Sharifa Minkabo\\   Department of Mathematics and Statistics\\
       College of Engineering and Science\\
		Louisiana Tech University\\
                    Ruston, LA USA\\
        Tel: +1 318 257 3529 ;  Fax: +1 318 257 2182\\
         \texttt{dcahoy@latech.edu}}

\allowdisplaybreaks

\usepackage[sort]{natbib} 
\usepackage[plainpages=false,pdfpagelabels,hypertexnames,linktocpage, colorlinks, citecolor=red,linkcolor=blue]{hyperref}

\begin{document}

\setlength{\parindent}{0pt}
\setlength{\parskip}{2pt}

\maketitle


\begin{abstract}
\indent

We propose  point  estimators  for the three-parameter (location, scale, and the fractional parameter) variant distributions  generated by a Wright function.  We also provide uncertainty quantification procedures  for the  proposed point estimators  under certain conditions. The  class of densities  includes the three-parameter one-sided and the three-parameter symmetric bimodal  $M$-Wright family of distributions.  The one-sided family  naturally generalizes  the  Airy and  half-normal models.   The   symmetric class  includes   the symmetric Airy and  normal or  Gaussian densities.  The  proposed interval estimator for the scale parameter outperformed the estimator derived in \cite{cah12} when the location parameter is zero.   We obtain  the asymptotic covariance structure for the scale and fractional parameter estimators, which allows estimation of the correlation. The coverage probabilities  of the interval estimators   slightly depend on the proposed  location parameter estimators. For the symmetric case,  the sample mean (or median) is favored than the median (or mean) when the fractional parameter is greater (or lesser) than 0.39106 in terms of their asymptotic relative efficiency.  The estimation algorithms  were tested using synthetic data and were compared with their bootstrap counterparts. The proposed inference procedures  were demonstrated on age and height data.

\vspace{0.1in}


\end{abstract}

\section{Introduction}

The $M$-Wright function has been increasingly gaining popularity from  several areas of study particularly in mathematics, engineering and physics.  It is often a probability density function in space which  solves time-fractional diffusion processes   \citep[see][]{mtm08}. As a solution, the $M$-Wright density naturally models the increments or  the 'space' component of the above processes at any given time.   It is also  used as a subordinator (as the operational time rather than the physical time) for time-fractional differential equations \citep{pas14}, for a  multi-point probability model  of the generalized grey Brownian motion that includes the well-known standard and fractional Brownian motions, and for pure linear birth processes \citep[see][]{bao10,cap12}. The single-parameter  positive-sided $M$-Wright function  takes the following  form:
\begin{equation}
M_\alpha (x) =\sum\limits_{j=0}^\infty \frac{ (-x)^j}{j! \Gamma [-\alpha j + (1-\alpha) ]}  =  \frac{1}{\pi} \sum\limits_{j=1}^\infty \frac{ (-x)^{j-1}}{(j-1)!} \Gamma (\alpha j) \sin ( \pi \alpha j)   \label{MW}
\end{equation}
where $x \in \mathbb{R}^+$, and $0 < \alpha < 1$ is the fractional parameter.  The last equality in the preceding equation follows from the reflection formula for the gamma function $\Gamma (1 - \alpha (j+1)) = \frac{\pi}{\Gamma ( \alpha (j+1) ) \sin (\pi \alpha (j+1) ) }$ and transformation $ j+1 \to j.$   We have the exponential  density $(\alpha=0^+)$ as a limiting case and the    Airy ($\alpha=1/3$) and  half-normal ($\alpha=1/2$)  \citep[see][]{mmp10} distributions as special cases where
\begin{equation}
M_{1/2}(x)=\frac{1}{\sqrt{\pi}}e^{-x^2/4}.
\end{equation}
Moreover, 
\[
M_\alpha(0^+) = 1/ \Gamma (1 - \alpha), \quad \text{and} \quad  M_{1^-}(x) =  \delta (x-1),
\]
where $\delta (\cdot)$ is the generalized Dirac function.   The Laplace transform of (\ref{MW}) is
\begin{equation}
 \mathbf{E}  \left(  e^{-  \beta X} \right)  = \phi_X(\beta)=  \sum\limits_{j=1}^\infty \frac{\left( -\beta \right)^j}{j! \;  \Gamma (1+ \alpha j )}
\end{equation}
which  is the Mittag-Leffler function.  The positive-sided $M$-Wright random variable has the structural representation
\begin{equation}
X \stackrel{d}{=} S^{-\alpha},
\end{equation}
where  $S$ follows an $\alpha^+$-stable distribution \citep{zol86}  with $\phi_S (\beta) =\exp (-\beta^\alpha)$.  The $\kappa$th moment   \citep[see][]{psw05} is known to be
\begin{equation}
\mathbf{E} X^\kappa = \frac{ \Gamma (1 + \kappa)}{\Gamma (1 + \alpha \kappa)}, \qquad \kappa>-1, 
\end{equation}
giving the mean and variance as 
\begin{equation}
\mu_x= \frac{1}{\alpha\Gamma (\alpha  )}, \quad\text{and} \quad  \sigma_x^2 = \frac{1}{\alpha \Gamma (2\alpha)} - \frac{1}{\left( \alpha \Gamma (\alpha)\right)^2},
\end{equation}
correspondingly. The coefficient of variation is straightforward to calculate as
\begin{equation}
 \frac{\sigma_x}{\mu_x} = \sqrt{\frac{2 \alpha \Gamma (\alpha) \Gamma (1 + \alpha)}{\Gamma (1 + 2\alpha)}-1} \;\; = \; \;\begin{cases} 1  \; &\;  \alpha=0, \\
\sqrt{\pi/2 -1}  \; &\;   \alpha=1/2,\\
0 \; & \;  \alpha=1.
\end{cases}  
\end{equation}
 The rest of the paper is organized as follows. The   one-sided $M$-Wright density, its properties, and test results are presented in Section 2.  The extension to the symmetric case are in Section 3.   The applications and concluding remarks are given in Sections 4 and 5, respectively.

\section{One-sided   \textit{M}-Wright distribution}

The  three-parameter one-sided $M$-Wright density function  has the following form:
\begin{equation}
M_{\alpha, \rho, \mu} (x) = \frac{1}{\rho} M_\alpha \left( \frac{x-\mu}{\rho} \right),   \qquad x>\mu,  
\end{equation}
where $\mu\in \mathbb{R}$ and $\rho \in \mathbb{R}^+$ are the shift and scale parameters, respectively.   Below are some forms of the densities in this family.

\begin{figure}[h!t!b!p!]
     \centering
			 \includegraphics[height=3in, width=4.5in]{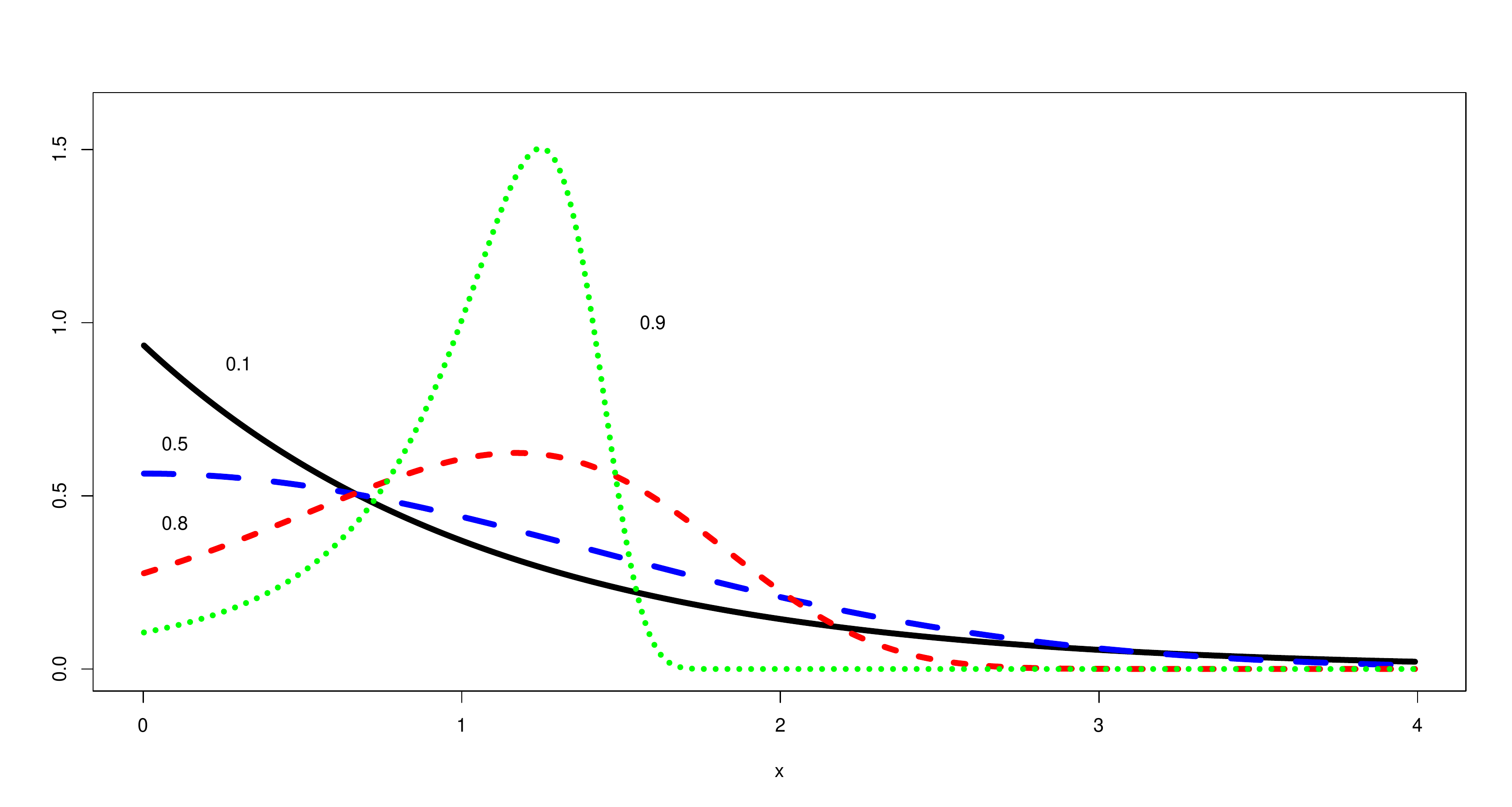}
       \renewcommand\abovecaptionskip{0pt}
       \caption{\emph{The one-sided $M$-Wright density  for  $\alpha=0.1, 0.5, 0.8, 0.9;  \rho=1,  \; \mu=0.$}}
     \renewcommand\belowcaptionskip{0pt}
\label{f0}
\end{figure}

\noindent\textbf{Case 2.1: $\bm{\mu=0}$}

If 
$
X\stackrel{d}{=} M_{\alpha, \rho, 0} (x) 
$
then
\begin{equation}
\hspace{-0.3in} \mathbf{E} X^\kappa = \frac{\rho^\kappa   \Gamma (1 + \kappa)}{\Gamma (1 + \alpha \kappa)},  \;\; \phi_X(\beta)= E_\alpha (-\beta \rho),   \;\; \phi_\rho (\beta)= \beta^{1-1/\alpha} e^{-x\beta},  \;\; \text{and} \;\;X\stackrel{d}{=}\rho S^{-\alpha}.    \label{2.1}
\end{equation}

Given $X_1, X_2, \ldots, X_n \; \;  \stackrel{iid}{=} \;  M_{\alpha, \rho, 0} (x),$   and applying the log transformation  to the absolute value of the random variable $X$ given in (\ref{2.1}), we obtain
\begin{equation}
X^{'} \stackrel{d}{=} \log (\rho ) -\alpha S^{'},  \label{2.2}  
\end{equation}
where $X^{'}= \log (|X|)$, and $S^{'}=\log (S)$.   From \cite{cah12},  the mean and variance are
\begin{equation}
\mu_{X^{'}}= \log (\rho ) + \gamma (\alpha-1), \quad \text{and} \quad \sigma_{X^{'}}^2= \frac{\pi^2}{6} \left(1-\alpha^2 \right),  \label{2.4} 
\end{equation}
respectively,  where  $\gamma \approx 0.5772156649$ is the Euler's constant.  Moreover,  the following point estimators of $\alpha$ and $\rho$ are obtained:
\begin{equation}
\widehat{\alpha}=\sqrt{1-\frac{6\hat{\sigma}_{X^{'}}^2}{\pi^2}},   \quad  \text{and}  \quad \widehat{\rho}= \exp \left( \widehat{\mu}_{X^{'}} + \gamma (1-\widehat{\alpha}) \right). \label{2.5}  
\end{equation}

\emph{Proposition 1. Let $X_1, X_2, \ldots, X_n \; \;  \stackrel{iid}{=} \; M_{\alpha, \rho, 0} (x) $. Then}
\begin{equation}
\sqrt{n}\left( \begin{array}{c}  \widehat{\alpha} - \alpha  \\  \widehat{\rho} - \rho  \end{array} \right) \stackrel{d}{\longrightarrow} \textsl{N}\left( \bm{0}\; ,\;\bm{\Sigma'}\right),  \qquad  n \to \infty,
\end{equation} 
\emph{where }
 \begin{equation}
\bm{\Sigma'} =  \left(
       \begin{array}{cc}
          \sigma_{\widehat{\alpha} \widehat{\alpha}} &  \sigma_{\widehat{\alpha} \widehat{\rho}} \\
           \sigma_{\widehat{\alpha} \widehat{\rho}} &   \sigma_{\widehat{\rho} \widehat{\rho}} \\
       \end{array}
     \right), \qquad \sigma_{\widehat{\alpha}\widehat{\alpha}}  = \frac{11- \alpha^4}{10   \alpha^2} -1,
\end{equation}
 
\begin{equation}
\sigma_{\widehat{\alpha} \widehat{\rho}} = \frac {\rho (10  \alpha^2 -11 + \alpha^4)  \gamma - \left[ 60 \alpha (\alpha^3 -1) \zeta (3) \right] / \pi^2} {10 \alpha^2},   
\end{equation}

\begin{equation}
\sigma_{\widehat{\rho} \widehat{\rho}}= \frac{\rho^2\left(\; 360 \alpha(\alpha^3 -1)\gamma \zeta (3)  - ( \alpha^2 -1) \pi^2 (3(11 + \alpha^2) \gamma^2 + 5 \alpha^2 \pi^2 ) \; \right)  } {30   \alpha^2 \pi^2 },
\end{equation}

\emph{and $ \zeta (\cdot )$ is the Riemman zeta function.}

\emph{Proof.} Recall the  following key results in \cite{cah12}:  Let $\mu_j^{'}=\mathbf{E} \left( X^{'} - \mu_{X^{'}}\right)^j, \; j=3,4.$   Then the third and fourth central moments   are 
\begin{equation}
\mu_3^{'}= 2(\alpha^3-1)\zeta (3)  \quad \text{and}  \quad 
\mu_4^{'}= \frac{\pi^4(\alpha^4-10 \alpha^2 + 9)}{60}, \label{mu34} 
\end{equation}
respectively.  In addition,  if $
\widehat{\mu}_{X^{'}}= \overline{X^{'}} = \sum \limits_{j=1}^n X_j^{'}\big/  n \quad
\text{and}\quad \widehat{\sigma}_{X^{'}}^2= \sum
\limits_{j=1}^n \left(X_j^{'}-\overline{X^{'}}\right)^2\big/n  $
then  it is widely known that 
\begin{equation}
\sqrt{n}\left(
  \begin{array}{c}
    \widehat{\mu}_{X^{'}}-\mu_{X^{'}} \\
    \widehat{\sigma}_{X^{'}}^2 - \sigma_{X^{'}}^2  \\ \label{clt1}
  \end{array}
\right) \stackrel{d}{\longrightarrow}  \textsl{N} \left[\bm{0} , \bf{\Sigma} \right]
\end{equation}
as $n \to \infty$, where the variance-covariance matrix $\bf{\Sigma}$ is defined as
\begin{equation}
\bf{\Sigma} = \left(
       \begin{array}{cc}
         \sigma_{X^{'}}^2 & \mu_3^{'} \\
         \mu_3^{'} & \mu_4^{'}-\sigma_{X^{'}}^4 \\
       \end{array}
     \right),  
\end{equation} 
$\mu_3^{'}, \mu_4^{'}$, and $\sigma_{X^{'}}^2$ are given in (\ref{2.4}) and (\ref{mu34}). Using result (\ref{clt1}) and   the multivariate delta method, 
\begin{equation}
\sqrt{n}\big(\textbf{g}(\widehat{\bm{\theta}}_n)-\textbf{g}(\bm{\theta})\big)\stackrel{d}{\to} \textsl{N}\left( \bm{0},\; \bm{\dot{\textbf{g}}}(\bm{\theta})^{\text{T}}\bf{\Sigma}\bf{\dot{g}}(\bm{\theta})\right), 
\end{equation} 
where  $\widehat{\bm{\theta}}_n=(\widehat{\mu}_{X^{'}},  \widehat{\sigma}_{X^{'}}^2)^\text{T},  \bf{g}$ is a continuous mapping from $\mathbb{R}^2 \to\mathbb{R}^2$ given as 
\[
 \textbf{g}(\mu_{X^{'}},\sigma_{X^{'}}^2) \; = \; \left(  \; \sqrt{1- \frac{6 \sigma_{X^{'}}^2}{\pi^2}}  \quad, \quad   \exp \left( \mu_{X^{'}}+\gamma(1-\alpha) \; \right)  \right)^{\text{T}}
\]
and   $\bm{\dot{\textbf{g}}}(\bm{\theta})= \nabla \textbf{g}(\bm{\theta})^{\text{T}}$ is the gradient matrix given by 

\[
\bm{\dot{\textbf{g}}}(\mu_{X^{'}},\sigma_{X^{'}}^2)  = 
\]
\begin{equation}
\left(
                                         \begin{array}{cc}
                                           0 &  \exp \left( \mu_{X^{'}}+\gamma(1-\alpha)
\right)\\
& \\
                                             -3 \Big/ \left(  \pi^2  \sqrt{1- \frac{6 \sigma_{X^{'}}^2}{\pi^2}} \; \right) &  \; \; 
   \Big (3 \gamma \exp (\mu_{X^{'}}+\gamma)\sqrt{1- \frac{6 \sigma_{X^{'}}^2}{\pi^2}} \; \Big ) /  \left(  \pi^2  \sqrt{1- \frac{6 \sigma_{X^{'}}^2}{\pi^2}}\; \right)\\
                                         \end{array}
                                        \right). \qed  \label{e21}
\end{equation}

Note that  the covariance structure of the scale and fractional parameter estimators  given by $\sigma_{\widehat{\alpha} \widehat{\rho}} $ above allows estimation of the correlation. 

\emph{Corollary 1.} \emph{Let $X_1, X_2, \ldots, X_n \; \;  \stackrel{iid}{=} \; X$.  The $(1-\nu)100\%$ confidence intervals for  $\alpha$ and $\rho$ can   be approximated as}
\begin{equation}
\widehat{\alpha} \;  \pm \; z_{\nu/2}\sqrt{\frac{\left[\left(11-\widehat{\alpha}^4 \right)/\left(10  \widehat{\alpha}^2\right)\right] -1}{n}},
\end{equation}
\emph{and}

\begin{equation}
\widehat{\rho} \; \pm \; z_{\nu/2}\sqrt{\frac{\widehat{\rho}^2 ( 360\widehat{\alpha}(\widehat{\alpha}^3-1 )\gamma \zeta (3)  -(\widehat{\alpha}^2-1) \pi^2 (3(11 +\widehat{\alpha}^2) \gamma^2 + 5 \widehat{\alpha}^2 \pi^2)  )}{30 n  \widehat{\alpha}^2 \pi^2}},
\end{equation}
\emph{correspondingly, where $z_{\nu/2}$ is the $(1-\nu/2)$th quantile of the standard normal distribution, and $0 <\nu<1$.}

\emph{Proof.} Immediately follows from Proposition 1  and is omitted.   \qed

We tested our estimators by simulating the bias ($100| \widehat{\theta} - \theta)|/\theta$), the median absolute deviation (MAD), and   the $95\%$ coverage probabilities for the proposed methods and the bootstrap percentile counterparts(with '*') corresponding to several parameter combinations.  Table 1 suggests that bias  is as large as $5\%$  and as little as $0.21\%$   when $n=10^4$.  Reduction in variability is also apparent  as the sample size goes large. It can be seen that the smaller the parameter $\alpha$, the slower the reduction in variability and bias  regardless of the sample size. Nevertheless, we conclude that these point estimators are consistent and asymptotically unbiased. Table 2  reveals that the proposed interval estimator of  the scale parameter quickly captured (e.g,  $n=100$ and $\alpha =0.6$)  the true nominal level than the one in \cite{cah12} as the sample size goes large.   Furthermore, Table 2 illustrates that the large-sample interval estimator outperformed the percentile bootstrap method for estimating $\alpha$ especially when $n \leq 1000$. Note that the large-sample formula   is  faster to calculate than the resampling-based method especially for large sample sizes.

\begin{table}[h!t!b!p!]
\caption{\emph{Mean estimates of and  dispersions from the true parameters  $\alpha$, and $\rho$. }} 
\centerline {
\begin{small}
\begin{tabular*}{5.8in}{@{\extracolsep{\fill}}c||c|cc|cc|cc}
\multirow{2}{*}{$(\alpha, \rho)$} &  &  \multicolumn{2}{c|}{$n=100$}  &  \multicolumn{2}{c|}{$n=1000$} &  \multicolumn{2}{c}{$n=10000$}  \\
 & &   $\%$ Bias &   $\%$ MAD&    $\%$ Bias &   $\%$ MAD &   $\%$ Bias &   $\%$ MAD \\
\hline \hline
\multirow{2}{*}{$(0.4, 150)$}  
&  $\widehat{\alpha}$ & 32.796 & 31.795 & 16.143  & 19.346 & 4.706 & 5.922\\
&  $\widehat{\rho}$ & 7.899 & 9.806 &  3.266 & 3.854 & 0.936 & 1.158\\                                    
\hline
\multirow{2}{*}{$(0.6, 8.77)$}  
&  $\widehat{\alpha}$ & 17.443 & 19.959 & 5.852 & 7.176 & 1.887 & 2.325\\
&  $\widehat{\rho}$ & 6.437 & 7.648 & 1.955 & 2.435 &  0.641 & 0.809\\
\hline
\multirow{2}{*}{$(0.8, 375)$}  
&  $\widehat{\alpha}$& 7.842 & 9.505 & 2.513 & 3.051 & 0.763 & 0.935\\
 &  $\widehat{\rho}$ & 4.423 & 5.273 & 1.351 & 1.706 &  0.408 & 0.502\\
\hline
\multirow{2}{*}{$(0.95, 1000)$}  
&  $\widehat{\alpha}$& 2.871 & 2.686 & 0.924 & 1.147 & 0.294 & 0.359\\
 &  $\widehat{\rho}$ &  2.329 & 2.769 & 0.687 & 0.838 & 0.213 & 0.268\\                            
\end{tabular*}
\end{small}
}
  \label{t11}
\end{table}


\begin{table}[h!t!b!p!]
\caption{\emph{Coverage probabilities   of 95\% interval estimates   for different values of $\alpha$, and $\rho$.}}
 \begin{small}
 \centerline {
\begin{tabular*}{3.7in}{c||c|c|c|c}
 $(\alpha, \rho)$ &  &  $n=100$  & $n=1000$ &  $n=10000$  \\
  \hline \hline
\multirow{6}{*}{$(0.4, 150)$}  
              &  $\widehat{\alpha}$  & 0.917  & 0.952  & 0.956 \\
              &  $\widehat{\rho}$  & 0.941 & 0.957  & 0.949 \\
              &  $\widehat{\alpha*}$  & 0.884 & 0.928 & 0.950\\
              &  $\widehat{\rho*}$ & 0.942 & 0.941 & 0.947\\
\hline
\hline
\multirow{6}{*}{$(0.6, 8.77)$}  &  $\widehat{\alpha}$   &  0.950 & 0.950 & 0.953\\
&  $\widehat{\rho}$     & 0.949 & 0.958 & 0.951\\
&  $\widehat{\alpha*}$  &  0.873 & 0.931 & 0.944\\
&  $\widehat{\rho*}$ & 0.943  & 0.954 & 0.950\\
\hline
\multirow{6}{*}{$(0.8, 375)$} 
              &  $\widehat{\alpha}$   & 0.964 & 0.950 & 0.954\\
              &  $\widehat{\rho}$  &   0.942 &  0.958 & 0.952\\
              &  $\widehat{\alpha*}$  & 0.831 & 0.925 & 0.948\\
              &  $\widehat{\rho*}$ &  0.935 & 0.954 & 0.948\\
 \hline
\multirow{6}{*}{$(0.95, 1000)$}  
&  $\widehat{\alpha}$  & 0.960 & 0.922 & 0.952\\ 
&  $\widehat{\rho}$   & 0.902 & 0.948 & 0.950\\
&  $\widehat{\alpha*}$  & 0.724 & 0.888 & 0.931\\
&  $\widehat{\rho*}$ & 0.922 & 0.948 & 0.947\\ 
\end{tabular*}
}
\end{small}
  \label{t33}
\end{table}

\noindent\textbf{Case 2.2: $\bm{\mu\neq 0}$}

Consider the location-scale structure
\begin{equation}
X \stackrel{d}{=} \mu + \rho S^{-\alpha},  \quad \text{and}\quad X>\mu. \label{1mw}
\end{equation}

\emph{Proposition 2.} \emph{Let $X_1, X_2, \ldots, X_n \; \;  \stackrel{iid}{=} \; X$  in (\ref{1mw}).   A   $(1-\nu)100\%$ confidence interval for the shift parameter $\mu$ is}
\begin{equation}
 \left(  \widehat{\mu}- q_\nu \widehat{\rho}  \; \; , \; \;  \widehat{\mu}  \right),
\end{equation}
\emph{where $q_{\nu}>0$ is the  $(1-\nu^{1/n})$th  quantile of   $M_{\alpha,1,0}$ and  $\widehat{\mu}=\min\{X_i \}_{i=1}^n$}.  

\emph{Proof.} Note that  
\begin{equation}
P(\mu < \widehat\mu < \mu +  q_{\nu} \rho) = 1-\nu,
\end{equation}
which suggests that 
\begin{eqnarray}
P( \widehat{\mu} >\mu + q_{\nu}\rho ) &=& \nu \\
&=& \prod\limits_{i=1}^n P(X_i > \mu + q_{\nu} \rho  )\\
&=&  \prod\limits_{i=1}^n P(\mu + \rho S_i^{-\alpha} > \mu + q_{\nu}\rho  )\\
&\implies&  P(S^{-\alpha} > q_{\nu}) = \nu^{1/n}.\qed
\end{eqnarray}

For reproducibility, we estimate   $q_{\nu} $ by generating $10^6$  random variates from    $M_{\widehat \alpha,1,0}$ and  use the approximately median-unbiased  (type 8 of the \texttt{quantile} function  in R) estimator  to calculate the $(1-\nu^{1/n})$th quantile as recommended by \cite{haf96}.  Note also that we directly use the point estimators  obtained in  case 2.1  after subtracting $\widehat{\mu}$  from the observed data.

Upon testing, Table 3  generally indicates  similar observations and conclusions about the estimators of $\alpha$ and $\rho$  as in Table 1.  The mean and dispersion  of $\widehat \mu$ seem  to be   large  when $\nu \approx 1$. Overall, the proposed point estimators are consistent.  In addition, Table 4 shows that the proposed    interval  estimator for $\mu$ seems to capture the true nominal rate even when the sample size is as small as 100 with $\alpha << 1$.    Comparing Tables 3 and 4 with Tables 1 and 2, correspondingly,  reveals  that the variability induced by the subtraction of the minimum from the data  does not seem to seriously  affect the performance of the proposed estimators.


\begin{table}[h!t!b!p!]
\begin{small}
\caption{\emph{Mean estimates of and  dispersions from the parameters  $\alpha$, $\rho$, and $\mu$.}} \centerline {
\begin{tabular*}{5.8in}{@{\extracolsep{\fill}}c||c|cc|cc|cc}
\multirow{2}{*}{$(\alpha, \rho, \mu)$} &  &  \multicolumn{2}{c|}{$n=100$}  &  \multicolumn{2}{c|}{$n=1000$} &  \multicolumn{2}{c}{$n=10000$}  \\
 & &   $\%$ Bias &   $\%$ MAD&    $\%$ Bias &   $\%$ MAD &   $\%$ Bias &   $\%$ MAD \\
  \hline \hline
\multirow{2}{*}{$(0.4, 150, -78)$}  
&  $\widehat{\alpha}$ & 32.459 & 31.174  & 16.026 & 19.658  &  4.919 &  6.116\\
&  $\widehat{\rho}$ & 7.706 & 9.732  & 3.060 & 3.559  & 0.949 &  1.201\\
&  $\widehat{\mu} $ & 2.032 & 2.055  & 0.220 & 0.221 &  0.022 & 0.023\\
\hline
\multirow{2}{*}{$(0.6, 8.77, 25.2)$}  
&  $\widehat{\alpha}$ & 17.258 & 18.863 & 6.017 & 7.427 &  1.871 &  2.358\\
&  $\widehat{\rho}$ & 6.737 & 8.896 & 2.012 & 2.487 & 0.613 &  0.761\\
 &  $\widehat{\mu} $ & 0.587 & 0.599 &  0.060 &  0.056  & 0.006 &  0.006\\
\hline
\multirow{2}{*}{$(0.8, 375, 375)$}  
&  $\widehat{\alpha}$& 7.597 & 8.491 & 2.716 & 3.407 & 0.821 & 1.051\\
&  $\widehat{\rho}$ & 5.092 & 5.777 &  1.374 & 1.681  & 0.429 & 0.550\\
&  $\widehat{\mu} $ & 3.521 & 3.404 & 0.333 & 0.331  & 0.036 & 0.036\\
\hline
\multirow{2}{*}{$(0.95, 1000,500)$}  
&  $\widehat{\alpha}$& 2.971 & 2.926 &  0.940 & 1.147 &  0.309 & 0.397\\
&  $\widehat{\rho}$ & 13.693 & 11.921  & 1.946 & 1.710 &  0.297 & 0.317\\
&  $\widehat{\mu} $ & 6.936 &  6.890  & 2.764 & 2.690 &  0.296 & 0.278\\
\end{tabular*}
}
  \label{t111}
\end{small}
\end{table}

\begin{table}[h!t!b!p!]
\caption{\emph{Coverage probabilities of 95\% interval estimates  for different values of $\alpha$, $\rho$, and   $\mu$.}}
 \begin{small}
 \centerline {
\begin{tabular*}{4.5in}{@{\extracolsep{\fill}}c||c|c|c|c}
 $(\alpha, \rho, \mu)$ &  &  $n=100$  & $n=1000$ &  $n=10000$  \\
  \hline \hline
\multirow{6}{*}{$(0.4, 150, -78)$}  
&  $\widehat{\alpha}$  &  0.921&  0.945 & 0.947\\
 &  $\widehat{\rho}$  &  0.944 &  0.961 &  0.956\\
&  $\widehat{\alpha*}$  &  0.951  & 0.953  & 0.947\\
&  $\widehat{\rho*}$ & 0.952  & 0.967 & 0.954\\
& $\widehat{\mu}$ & 0.962 &  0.951 & 0.945\\ 
\hline
\multirow{6}{*}{$(0.6, 8.77, 25.2)$}  
&  $\widehat{\alpha}$   &  0.955 & 0.955 &  0.952\\
&  $\widehat{\rho}$     & 0.931 & 0.954 &  0.958\\
&  $\widehat{\alpha*}$  &  0.976  & 0.968  & 0.963\\
&  $\widehat{\rho*}$ & 0.940  & 0.956   & 0.958\\
& $\widehat{\mu}$ & 0.943 & 0.946  & 0.949\\
\hline
\multirow{6}{*}{$(0.8, 375, 375)$}  
&  $\widehat{\alpha}$   & 0.966 & 0.940 &  0.947\\
&  $\widehat{\rho}$  &   0.865  & 0.946 &  0.951\\
 &  $\widehat{\alpha*}$  & 0.980  & 0.971 &  0.953\\
&  $\widehat{\rho*}$ & 0.887 & 0.945  & 0.952\\
 & $\widehat{\mu}$ & 0.911 & 0.952 &  0.949\\ 
 \hline
\multirow{6}{*}{$(0.95, 1000,500)$}  
&  $\widehat{\alpha}$  & 0.963 &  0.947 &  0.949\\ 
&  $\widehat{\rho}$   & 0.945  & 0.957 & 0.951\\
&  $\widehat{\alpha*}$  & 0.996 &  0.963 &  0.966\\
&  $\widehat{\rho*}$ & 0.931&  0.959  & 0.934\\ 
& $\widehat{\mu}$ & 0.907 & 0.941 & 0.947\\
\end{tabular*}
}
\end{small}
  \label{t333}
\end{table}

\pagebreak
\section{Symmetric  \textit{M}-Wright distribution}

Replacing $x$ by $|x|$ in  (\ref{MW}) and dividing   (\ref{MW})  by two,  the three-parameter symmetric $M$-Wright density   can be written as 
 
\begin{equation}
M_{\alpha, \rho, \mu} (x) = \frac{1}{2\rho} M_\alpha \left( \frac{|x-\mu|}{\rho} \right),   \qquad x \in \mathbb{R},  
\end{equation}
where $\mu  \in \mathbb{R}$ and $\rho \in \mathbb{R}^+$ are the location and scale parameters, respectively.  The Laplace or double exponential $(\alpha=0^+)$ is a limiting case while the Gaussian or  normal ($\alpha=1/2$)  \citep[see][]{mmp10} distributions are special cases. Moreover,

\begin{equation}
X \stackrel{d}{=} \mu + \rho U S^{-\alpha}, \qquad U \stackrel{ind}{=} (1/2)[ \delta (u+1) + \delta(u-1)], \label{2mw}
\end{equation}
where `$ind$' means \emph{independent}.

\noindent\textbf{Case 3.1: $\bm{\mu= 0}$}

The $M$-Wright function in two variables  that  is centered at zero satisfies the following  transformation:

\[
\phi_{|X|}( - \beta) = 2 E_{2\alpha} \left(-\beta^2 \right).
\]

When  $\alpha=1/2$,  we get the Gaussian density
\begin{equation}
\frac{1}{2} M_{1/2, \rho} (-|y|) = \frac{1}{2 \sqrt{ \pi } \rho^{-1} } \exp \bigg(  \frac{-y^2}{4 \rho^{-2}}\bigg)
\end{equation}
with mean zero and variance $2\rho^{-2}$.   
It is easy to show that 
\begin{equation}
 \vert U S^{-\alpha}\vert \stackrel{d}{=}S^{-\alpha}.
\end{equation}
The preceding result allows us to estimate  the parameters of the two-sided symmetric $M$-Wright distribution  using the properties of its one-sided non-symmetric counterpart. Furthermore, the formula for the integer-order moments of the symmetric two-parameter $M$-Wright distribution centered at zero can be deduced as
\begin{equation}
\mathbf{E} X^\kappa  = \left\{
                           \begin{array}{ll}
                                 \frac{ \rho^\kappa  \Gamma (1 + \kappa)}{\Gamma (1 + \alpha \kappa)} & \text{if} \;j \; \text{is even}, \\
& \\
                              0 & \text{if} \;j \; \text{is odd}.
                           \end{array}
                         \right.
\end{equation}

\begin{figure}[h!t!b!p!]
     \centering
			 \includegraphics[height=3in, width=4in]{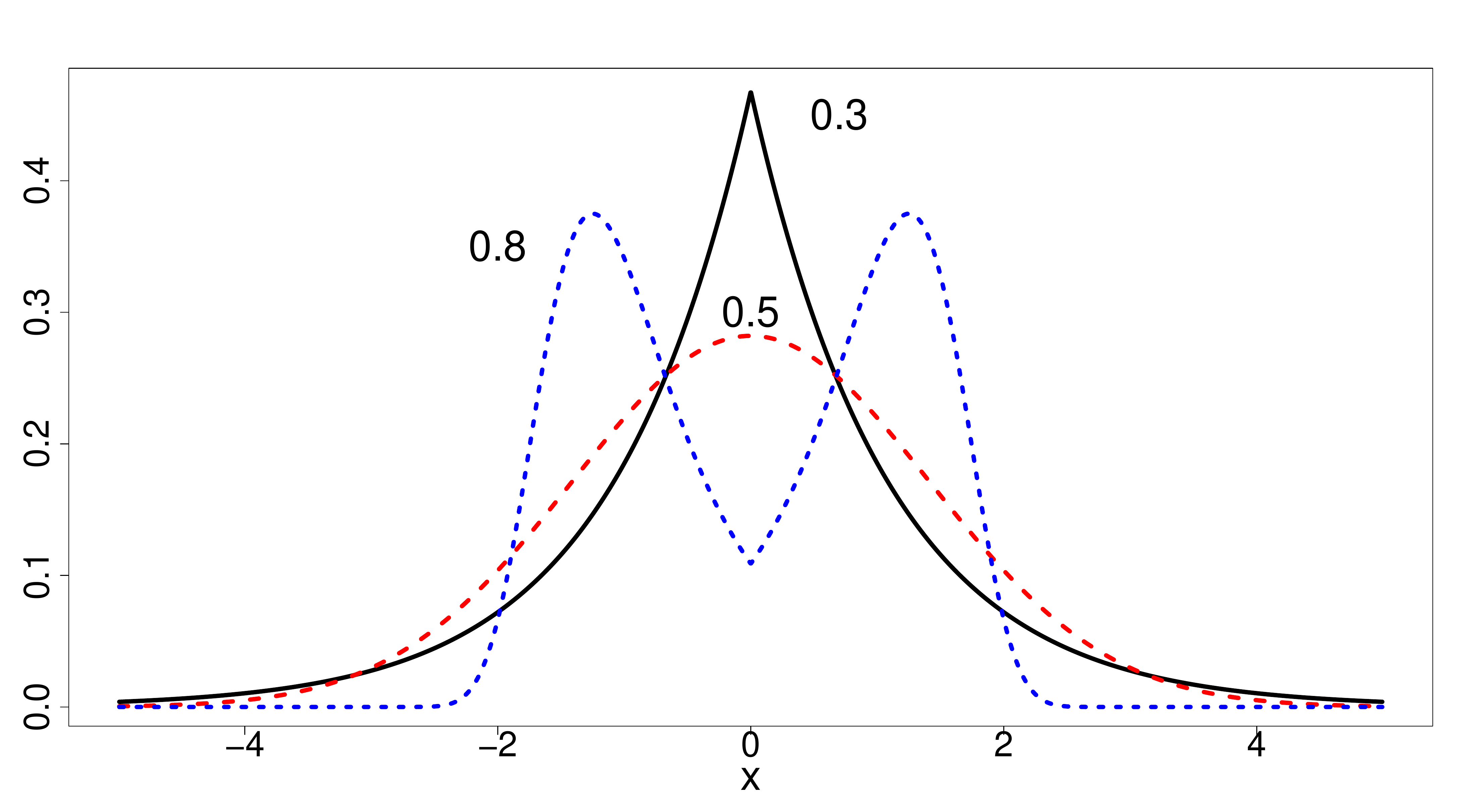}
       \renewcommand\abovecaptionskip{0pt}
       \caption{\emph{The symmetric  $M$-Wright density  for  $\alpha=0.3, 0.5, 0.8;  \rho=1,  \; \mu=0$.}}
     \renewcommand\belowcaptionskip{0pt}
\label{f1}
\end{figure}

For completeness, we reproduce Figure \ref{f1}  from \citep{cah12,cah12b} to emphasize the flexibility of the symmetric single-parameter  $M$-Wright density. 

\noindent\textbf{Case 3.2: $\bm{\mu\neq 0}$}

\emph{Proposition 3.} \emph{Let $X_1, X_2, \ldots, X_n \; \;  \stackrel{iid}{=} \; X$ in (\ref{2mw}).} Then   
\begin{equation}
\sqrt{n} \left(\overline{X} \;  - \;  \mu \right) \stackrel{d}{\longrightarrow} N\left(0, \;\frac{\rho^2}{\alpha \Gamma (2 \alpha ) }\right)
\end{equation}
\emph{and}
\begin{equation}
\sqrt{n} \left( \widetilde{X} \;  - \;  \mu \right) \stackrel{d}{\longrightarrow} N \left( 0, \;   \rho^2  \; \Gamma (1- \alpha )^2 \right),
\end{equation}
\emph{as $n \to \infty$ where $\widetilde{X}$ is the sample median.  }

\emph{Proof.} Directly follows from the standard large sample results for mean and median of random samples. \qed

Thus, the asymptotic relative efficiency of $\overline{X}$ to $\widetilde{X}$ $\left(ARE\left(\overline{X}, \widetilde{X}\right)\right)$ is
\begin{equation}
ARE\left(\overline{X}, \widetilde{X}\right) = \left( \alpha \Gamma (2 \alpha) \Gamma ( 1-\alpha)^2 \right)^{-1}.
\end{equation}

\begin{figure}[h!t!b!p!]
     \centering
			 \includegraphics[height=3in, width=3.5in]{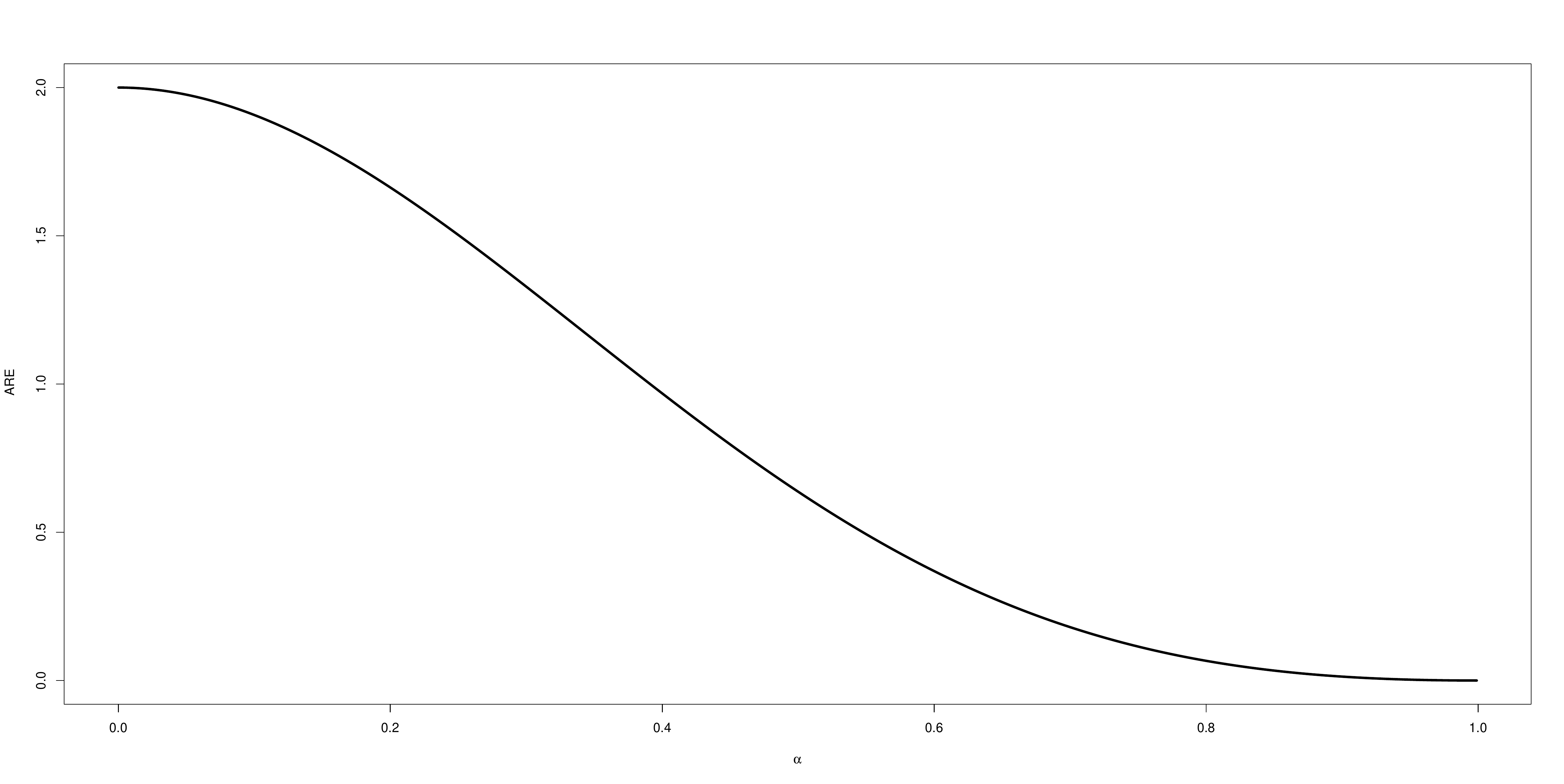}
       \renewcommand\abovecaptionskip{0pt}
       \caption{\emph{The asymptotic relative efficiency of $\overline{X}$ to $\widetilde{X}$ $\left(ARE\left(\overline{X}, \widetilde{X}\right)\right)$  as a function of $\alpha$.}}
     \renewcommand\belowcaptionskip{0pt}
\label{f2}
\end{figure}

Figure \ref{f2} displays the asymptotic relative efficiency of $\overline{X}$ to $\widetilde{X}$  as a function of $\alpha$.

\pagebreak
The relative efficiency above equals unity  if $\alpha=0.39106$.   Thus, the sample mean $\overline{X}$  is used for $\alpha>0.39106$. Otherwise, the sample median $\widetilde{X}$ is preferred when $\alpha<0.39106$ for relatively large samples.

\emph{Corollary 2.}  \emph{Let $X_1, X_2, \ldots, X_n \; \;  \stackrel{iid}{=} \; X$ in (\ref{2mw}).  From Proposition 3, the  approximate mean-based  $(1-\nu)\%$  confidence interval for $\mu$ is}
\begin{equation}
\overline{X} \; \pm \;  z_{\nu/2} \; \widehat{\rho} \; \left( \;  \widehat{\alpha} \; n\; \Gamma ( 2 \widehat{\alpha} ) \; \right)^{-1/2}
\end{equation}
\emph{while the approximate  median-based  $(1-\nu)\%$  confidence interval for $\mu$ is}
 \begin{equation}
\widetilde{X} \; \pm \; z_{\nu/2} \; \frac{\widehat{\rho}\; \Gamma (1- \widehat{\alpha})}{\sqrt{n}}.
\end{equation}

\emph{Proof.} Directly follows from the central limit theorem and the asymptotic normality of the sample median. \qed

Subtracting $\widehat{\mu}$ from the data and getting the absolute values  allow us to use the estimators of  $\alpha$ and $\rho$ from the  preceding section.

\newpage
For testing purposes, we used the sample mean as the location parameter estimator as $\alpha$ values are chosen to be at least 0.4.   Table 5 suggests  negligible increase (due to the variability induced by subtracting the mean from the data)  in both bias and MAD  for the proposed point estimators of  $\alpha$ and $\rho$ in comparison with Table 1 ($\mu=0$)  as  $n \to \infty$.

\begin{table}[h!t!b!p!]
\begin{small}
\caption{\emph{Mean estimates of and  dispersions from the true parameters  $\alpha$, $\rho$, and $\mu$.}} \centerline {
\begin{tabular*}{5.8in}{@{\extracolsep{\fill}}c||c|cc|cc|cc}
\multirow{2}{*}{$(\alpha, \rho, \mu)$} &  &  \multicolumn{2}{c|}{$n=100$}  &  \multicolumn{2}{c|}{$n=1000$} &  \multicolumn{2}{c}{$n=10000$}  \\
 & &   $\%$ Bias &   $\%$ MAD&    $\%$ Bias &   $\%$ MAD &   $\%$ Bias &   $\%$ MAD \\
  \hline \hline
\multirow{2}{*}{$(0.4, 150, -78)$}  
&  $\widehat{\alpha}$ & 33.282 & 31.708 & 15.402 & 16.742 & 4.841 & 6.076\\
&  $\widehat{\rho}$ & 7.821 & 9.771 &  2.928 & 3.365 & 0.958 & 1.233\\
&  $\widehat{\mu} $ & 21.865 & 26.410 & 6.842 & 8.610 & 2.237 & 2.868\\
\hline
\multirow{2}{*}{$(0.6, 8.77, 25.2)$}  
&  $\widehat{\alpha}$ & 17.082 & 20.509 & 6.114 & 7.422 &  1.959 & 2.536 \\
&  $\widehat{\rho}$ &  6.272 & 7.587 &  1.895 & 2.311 &  0.620 & 0.792\\
 &  $\widehat{\mu} $ &  3.691 & 4.536 &  1.136 & 1.389 &  0.387 & 0.494\\
\hline
\multirow{2}{*}{$(0.8, 375, 375)$}  
&  $\widehat{\alpha}$& 8.427 & 9.125 & 2.551 & 3.186 & 0.837 & 1.027\\
&  $\widehat{\rho}$ & 4.479 & 5.532 & 1.328 & 1.636 & 0.435 & 0.559\\
&  $\widehat{\mu} $ & 9.361 & 11.389  & 2.980 & 3.818 & 1.000 & 1.306\\
\hline
\multirow{2}{*}{$(0.95, 1000,500)$} 
&  $\widehat{\alpha}$&  2.841 & 2.822 & 0.958 & 1.192 & 0.310 & 0.382\\
&  $\widehat{\rho}$ &  2.536 & 2.927 &  0.695 & 0.885 & 0.217 & 0.271 \\
&  $\widehat{\mu} $ & 16.765 & 21.059 & 5.311 & 6.961 & 1.657 & 2.053\\
\end{tabular*}
}
\end{small}
\end{table}

\newpage
We also tested the proposed interval estimators  and compared with their bootstrap counterparts (using percentile method).  From Table 6, the large-sample interval estimator for $\alpha$ outperformed its  bootstrap counterpart especially when $n=100$.

\begin{table}[h!t!b!p!]
\caption{\emph{Coverage probabilities of 95\% interval estimates  for different values of $\alpha$, $\rho$, and  $\mu$.}}
 \begin{small}
 \centerline {
\begin{tabular*}{4in}{@{\extracolsep{\fill}}c||c|c|c|c}
 $(\alpha, \rho, \mu)$  &  &  $n=100$  & $n=1000$ &  $n=10000$  \\
 \hline 
\multirow{6}{*}{$(0.4, 150, -78)$}  
&  $\widehat{\alpha}$  &  0.906 & 0.956  & 0.951\\
&  $\widehat{\rho}$  & 0.958 & 0.960 &  0.957\\
& $\widehat{\mu}$ &   0.940  & 0.955 & 0.955\\ 
&  $\widehat{\alpha*}$  &  0.859 & 0.929 & 0.943\\
&  $\widehat{\rho*}$ & 0.939 & 0.957 &  0.947\\
& $\widehat{\mu*}$ & 0.946 & 0.955 & 0.958\\ 
\hline
\multirow{6}{*}{$(0.6, 8.77, 25.2)$}  
&  $\widehat{\alpha}$   &  0.956 & 0.955 & 0.953\\
&  $\widehat{\rho}$     & 0.942 & 0.947 & 0.955\\
& $\widehat{\mu}$ & 0.952 & 0.956 & 0.952\\
&  $\widehat{\alpha*}$  &  0.887 & 0.938  & 0.939\\
&  $\widehat{\rho*}$ & 0.962 & 0.948 & 0.955\\
& $\widehat{\mu*}$ & 0.95 & 0.949 & 0.954\\
\hline
\multirow{6}{*}{$(0.8, 375, 375)$}  
              &  $\widehat{\alpha}$   & 0.969 & 0.959 & 0.945\\
              &  $\widehat{\rho}$  &   0.924 & 0.954 & 0.958\\
              & $\widehat{\mu}$ &  0.940 & 0.953 & 0.945\\ 
              &  $\widehat{\alpha*}$  & 0.862 & 0.941 & 0.944\\
              &  $\widehat{\rho*}$ & 0.944 & 0.949 & 0.954\\
              & $\widehat{\mu*}$ & 0.939 & 0.948 & 0.943\\ 
 \hline
\multirow{6}{*}{$(0.95, 1000,500)$}  
&  $\widehat{\alpha}$  & 0.976 & 0.955 & 0.950\\ 
&  $\widehat{\rho}$   & 0.891 & 0.950 & 0.953\\
& $\widehat{\mu}$ & 0.943 & 0.954 & 0.953\\
&  $\widehat{\alpha*}$  & 0.781 & 0.908 & 0.942\\
&  $\widehat{\rho*}$ & 0.910 & 0.953 & 0.951\\ 
& $\widehat{\mu*}$ & 0.953 & 0.951 & 0.949\\
\end{tabular*}
}
\end{small}
  \label{t3}
\end{table}

  
\pagebreak
\section{Applications}
 
We apply our methods on two real datasets that are available online (used in some researches)  using the statistical software R.   R codes are also available upon request through \texttt{dcahoy@latech.edu}.

\subsection{Ages of Major League Baseball players}

We consider the ages (in years) of 826 Major League Baseball (MLB) players.  The data was downloaded from the Statistics Online Computational Resource (SOCR) database (see  \texttt{http://wiki.stat.ucla.edu/socr/index.php/SOCR$\_$Data$\_$\\ Dinov$\_$020108$\_$HeightsWeights}).    The   one-sided $M$-Wright fit to the  data  yields the point and interval estimates    in Table \ref{est1}.  The minimum age of these players tends to be around 25 years old. The confidence interval estimate of the fractional parameter excludes the exponential ($\alpha=0^+$)  and the Airy ($\alpha=1/3$)  distributions  but includes the half-normal ($\alpha=1/2$)  model.  Using the asymptotic bivariate results in Section 2, the correlation between $\widehat{\alpha}$ and $\widehat{\rho}$ can be easily estimated  as -0.989, which indicates a strong inverse linear relationship.

\begin{table}[h!t!b!p!]
\caption{\emph{Estimates  for  $\mu$,   $\alpha$, and $\rho$.}}
 \vspace{0.2in}
 \centerline {
\begin{tabular*}{4in}{@{\extracolsep{\fill}}c||c|c} 
Parameter & Point estimate  & 95\% Confidence interval \\
  \hline \hline 
& \\ 
$\mu$ & 25.020 &  (  24.960\;,\;  25.020 )\\
& \\
   $\alpha$ & 0.473 &  ( 0.338\;,\; 0.607 ) \\
  & \\
   $\rho$  & 4.390 &  ( 4.094\;,\; 4.686 )\\
& \\
 \hline
\end{tabular*}
}
  \label{est1}
\end{table}

The two-sample Kolmogorov-Smirnov method (using R) was also used to test the fits of  100 simulated data sets (of same size with the observed data) using the parameter estimates.  The average p-value (0.841) indicated a reasonably good  fit. The succeeding figure demonstrates the   $M$-Wright  fit to the SOCR MLB age data   with the maximum likelihood fits of \texttt{gamma(shape=1.2994, rate=0.2605)},  \texttt{Weibull(shape=1.2177, scale=5.3071)}    and \texttt{lognormal(meanlog = 1.1752, \\  sdlog=1.1292)} distributions. By visual inspection, the    one-sided $M$-Wright distribution seems to provide the best fit. The picture also suggests that the one-sided $M$-Wright had the flexibility to model data  populations  which have  an inflection point (e.g., $\alpha=0.5$: half-normal) with mode at the origin or minimum  and their variants  corresponding to $\alpha \approx 0.5.$  It can also be checked that at the origin, the height is   $M_{\widehat{\alpha}, \widehat{\rho}, \widehat{\mu}} (\widehat{\mu} )= (\widehat{\rho}\cdot  \Gamma (1-\widehat{\alpha}) )^{-1} =  0.1352.$

\begin{figure}[h!t!b!p!]
     \centering
			 \includegraphics[height=3.2in, width=4.3in]{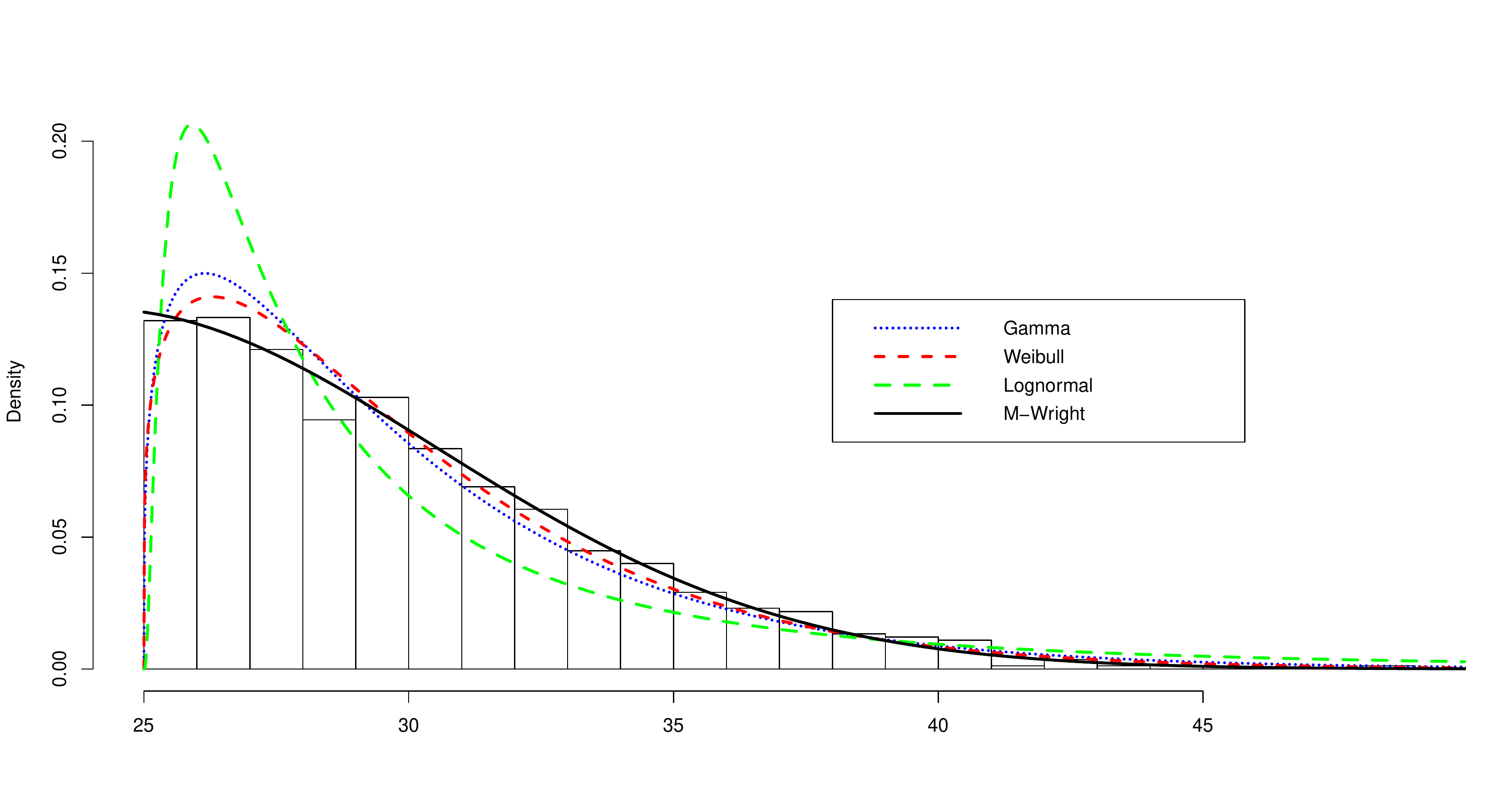}
 \caption{Model fits to ages of  MLB players.}
       \renewcommand\abovecaptionskip{0pt}
     \renewcommand\belowcaptionskip{0pt}
\end{figure}

\subsection{Human height and weight } 

 The dataset  contains 25000 records of human heights (in inches) and can be downloaded from the SOCR website. These data were obtained in 1993 by a Growth Survey of 25000 children from birth to 18 years of age recruited from Maternal and Child Health Centres (MCHC) and schools, and were used to develop Hong Kong's current growth charts for weight, height, weight-for-age, weight-for-height and body mass index (BMI). Below are the corresponding point and 95\% interval estimates  for the three parameters.  We used the sample mean as the point estimator as $\widehat{\alpha}$ is greater than the cutoff value of 0.39106 above. The interval estimate  seems not to favor the double-exponential $(\alpha=0^+)$  and  normal or Gaussian $(\alpha=0.5)$  densities to likely model the distribution of the children's heights.  The estimate of the correlation between $\widehat{\alpha}$ and $\widehat{\rho}$ is   -0.613, which indicates moderate negative association.

\begin{table}[h!t!b!p!]
\caption{\emph{Estimates  for   $\mu$, $\alpha$, and $\rho$.}}
 \vspace{0.2in}
 \centerline {
\begin{tabular*}{4in}{@{\extracolsep{\fill}}c||c|c} 
Parameter & Point estimate &   95\% Confidence interval  \\
  \hline \hline 
& \\ 
$\mu$ & 67.993 &   (  67.969\;,\;  68.017 )\\
& \\
   $\alpha$  & 0.481 &  ( 0.457\;,\; 0.505 ) \\
  & \\
   $\rho$  &1.352 &  ( 1.336\;,\; 1.369 )\\
& \\
 \hline
\end{tabular*}
}
  \label{ci1}
\end{table}

The two-sample Kolmogorov-Smirnov method (using R) was again used to test the fits of  100 simulated data sets (of same size with the observed data) using the parameter estimates above.  The average p-value (0.586) indicated a reasonably good  fit to the data. The following figure demonstrated the   fit   of the model to the SOCR height data.
 
\begin{figure}[h!t!b!p!]
     \centering
			 \includegraphics[height=3.2in, width=4.3in]{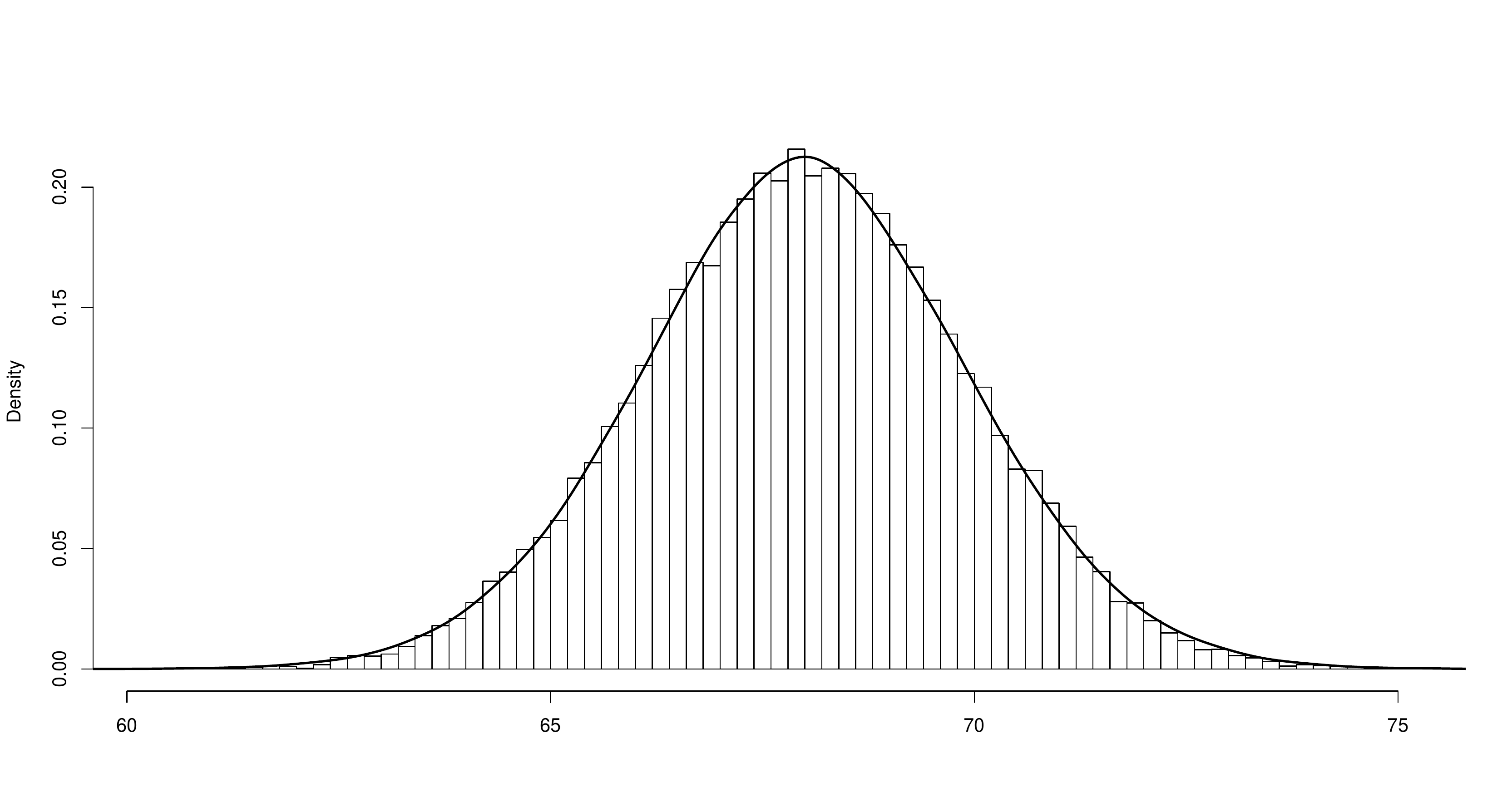}
 \caption{Symmetric $M$-Wright fit to 25,000 heights of children from birth to 18 years of age.}
       \renewcommand\abovecaptionskip{0pt}
     \renewcommand\belowcaptionskip{0pt}
\end{figure}


\section{Concluding Remarks}
Statistical inference  procedures for the  three-parameter $M$-Wright family of distributions were proposed.  The point  estimators of the location, scale and fractional parameters   were proven to be consistent and asymptotically unbiased.  The large-sample results allowed quantification of the uncertainty associated with the proposed point estimators.  The inference techniques were  also demonstrated using real data sets, which indicated the 'smoothing' effect of the   fractional parameter $\alpha \in (0, 1)$.  The proposed location parameter estimators did not  seriously affect the properties of the scale and fractional parameter estimates (point and interval).    The random number generation algorithms  were provided by the structural representations.  Improvements of these procedures using robust or Bayesian perspectives and the derivation of the trivariate  or joint asymptotic distribution of the location, scale, and fractional estimators   would be worth exploring in the future.

\section{Acknowledgment}
The authors are grateful to the  anonymous reviewers and co-editor-in-chief for their insightful comments and valuable suggestions that significantly improved the article. 

\section{References}
\renewcommand*{\refname}{}

\end{document}